\documentclass[twoside,twocolumn,9pt]{article}
\usepackage{extsizes}
\usepackage[super,sort&compress,comma]{natbib} 
\usepackage[version=3]{mhchem}
\usepackage[left=1.5cm, right=1.5cm, top=1.785cm, bottom=2.0cm]{geometry}
\usepackage{balance}
\usepackage{widetext}
\usepackage{times,mathptmx}
\usepackage{sectsty}
\usepackage{graphicx} 
\usepackage{lastpage}
\usepackage[format=plain,justification=justified,singlelinecheck=false,font={stretch=1.125,small,sf},labelfont=bf,labelsep=space]{caption}
\usepackage{float}
\usepackage{fancyhdr}
\usepackage{fnpos}
\usepackage[english]{babel}
\usepackage{array}
\usepackage{droidsans}
\usepackage{charter}
\usepackage[T1]{fontenc}
\usepackage[usenames,dvipsnames]{xcolor}
\usepackage{setspace}
\usepackage[compact]{titlesec}
\usepackage{soul}
\usepackage[hidelinks]{hyperref}
\usepackage{ifthen}
\usepackage[colorinlistoftodos,prependcaption,textsize=small]{todonotes}

\usepackage{epstopdf}

\definecolor{cream}{RGB}{222,217,201}
\usepackage{amsmath} 

\begin{document}

\pagestyle{fancy}
\thispagestyle{plain}
\fancypagestyle{plain}{

\fancyhead[C]{\includegraphics[width=18.5cm]{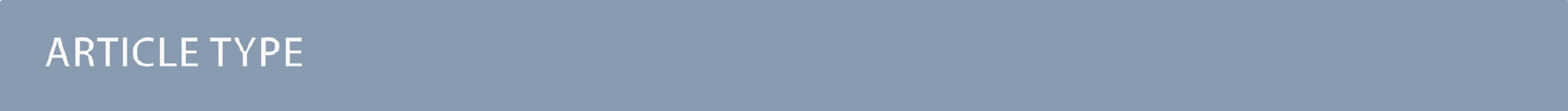}}
\fancyhead[L]{\hspace{0cm}\vspace{1.5cm}\includegraphics[height=30pt]{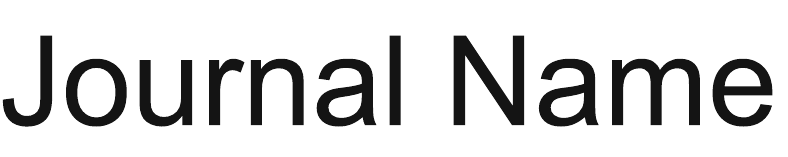}}
\fancyhead[R]{\hspace{0cm}\vspace{1.7cm}\includegraphics[height=55pt]{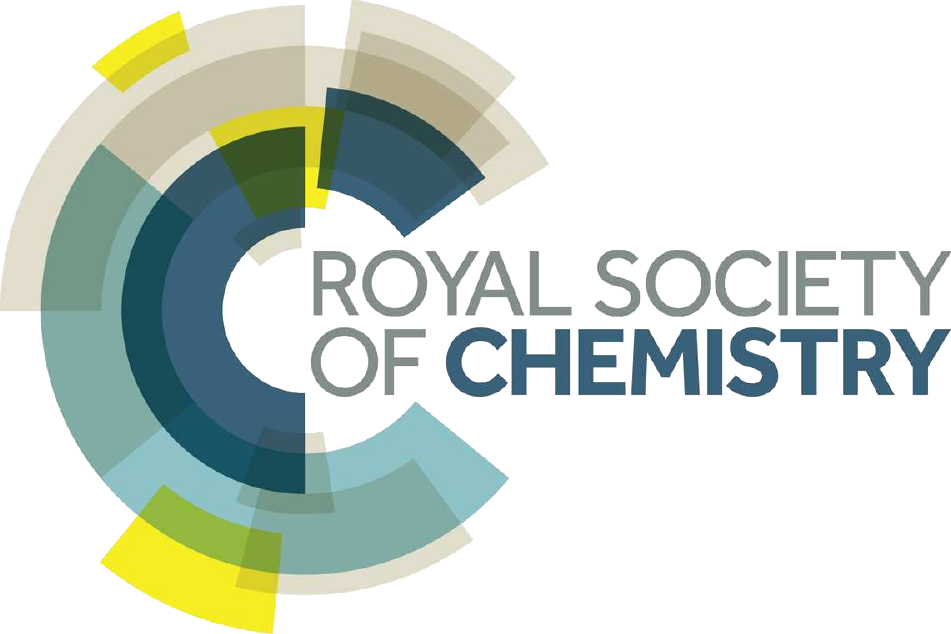}}
\renewcommand{\headrulewidth}{0pt}
}

\makeFNbottom
\makeatletter
\renewcommand\LARGE{\@setfontsize\LARGE{15pt}{17}}
\renewcommand\Large{\@setfontsize\Large{12pt}{14}}
\renewcommand\large{\@setfontsize\large{10pt}{12}}
\renewcommand\footnotesize{\@setfontsize\footnotesize{7pt}{10}}
\makeatother

\renewcommand{\thefootnote}{\fnsymbol{footnote}}
\renewcommand\footnoterule{\vspace*{1pt}%
\color{cream}\hrule width 3.5in height 0.4pt \color{black}\vspace*{5pt}} 
\setcounter{secnumdepth}{5}

\makeatletter 
\renewcommand\@biblabel[1]{#1}            
\renewcommand\@makefntext[1]%
{\noindent\makebox[0pt][r]{\@thefnmark\,}#1}
\makeatother 
\renewcommand{\figurename}{\small{Fig.}~}
\sectionfont{\sffamily\Large}
\subsectionfont{\normalsize}
\subsubsectionfont{\bf}
\setstretch{1.125} 
\setlength{\skip\footins}{0.8cm}
\setlength{\footnotesep}{0.25cm}
\setlength{\jot}{10pt}
\titlespacing*{\section}{0pt}{4pt}{4pt}
\titlespacing*{\subsection}{0pt}{15pt}{1pt}

\fancyfoot{}
\fancyfoot[LO,RE]{\vspace{-7.1pt}\includegraphics[height=9pt]{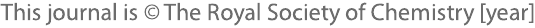}}
\fancyfoot[CO]{\vspace{-7.1pt}\hspace{13.2cm}\includegraphics{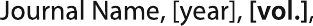}}
\fancyfoot[CE]{\vspace{-7.2pt}\hspace{-14.2cm}\includegraphics{RF}}
\fancyfoot[RO]{\footnotesize{\sffamily{1--\pageref{LastPage} ~\textbar  \hspace{2pt}\thepage}}}
\fancyfoot[LE]{\footnotesize{\sffamily{\thepage~\textbar\hspace{3.45cm} 1--\pageref{LastPage}}}}
\fancyhead{}
\renewcommand{\headrulewidth}{0pt} 
\renewcommand{\footrulewidth}{0pt}
\setlength{\arrayrulewidth}{1pt}
\setlength{\columnsep}{6.5mm}
\setlength\bibsep{1pt}

\makeatletter 
\newlength{\figrulesep} 
\setlength{\figrulesep}{0.5\textfloatsep} 

\newcommand{\topfigrule}{\vspace*{-1pt}%
\noindent{\color{cream}\rule[-\figrulesep]{\columnwidth}{1.5pt}} }

\newcommand{\botfigrule}{\vspace*{-2pt}%
\noindent{\color{cream}\rule[\figrulesep]{\columnwidth}{1.5pt}} }

\newcommand{\dblfigrule}{\vspace*{-1pt}%
\noindent{\color{cream}\rule[-\figrulesep]{\textwidth}{1.5pt}} }

\makeatother

\twocolumn[
  \begin{@twocolumnfalse}
\vspace{3cm}
\sffamily
\begin{tabular}{m{4.5cm} p{13.5cm} }

\includegraphics{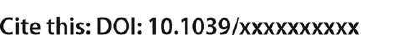} & \noindent\LARGE{\textbf{Topological Line Defects around Graphene Nanopores for DNA Sequencing$^\dag$}} \\
\vspace{0.3cm} & \vspace{0.3cm} \\

 & \noindent\large{Jariyanee Prasongkit,\textit{$^{a,b}$} Ernane F. Martins,\textit{$^{c,d}$}  F\'abio A. L. de Souza,\textit{$^{e}$}  Wanderl\~a L. Scopel,\textit{$^{f}$} Rodrigo G. Amorim,\textit{$^{g}$}} Vittaya Amornkitbamrung,\textit{$^{b,h}$}  Alexandre R. Rocha,\textit{$^{c}$} and Ralph H. Scheicher\textit{$^{d}$}   \\

\includegraphics{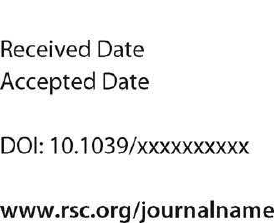} & \noindent\normalsize{
Topological line defects in graphene represent an ideal  way to produce highly controlled structures with reduced dimensionality that can be used in electronic devices.
In this work we propose using extended line defects in graphene to improve nucleobase selectivity in nanopore-based DNA sequencing devices. We use a combination of QM/MM and non-equilibrium Green's functions methods to investigate the conductance modulation, fully accounting for solvent effects. By sampling over a large number of different orientations generated from molecular dynamics simulations, we theoretically demonstrate that distinguishing between the four nucleobases using line defects in a graphene-based electronic device appears possible. The changes in  conductance are associated with transport across specific molecular states near the Fermi level and their coupling to the pore. Through the application of a specifically tuned gate voltage, such a device would be able to discriminate the four types of nucleobases more reliably than that of graphene sensors without topological line defects. 
}

\end{tabular}

 \end{@twocolumnfalse} \vspace{0.6cm}

  ]

\renewcommand*\rmdefault{bch}\normalfont\upshape
\rmfamily
\section*{}
\vspace{-1cm}


\footnotetext{\textit{$^{a}$~Division of Physics, Faculty of Science, Nakhon Phanom University, Nakhon Phanom, 48000, Thailand; E-mail: jariyanee.prasongkit@npu.ac.th}}
\footnotetext{\textit{$^{b}$~Thailand Center of Excellence in Physics, Commission on Higher Education, 328 Si Ayutthaya Road, Bangkok 10400, Thailand}}
\footnotetext{\textit{$^{c}$~Institute of Theoretical Physics, S\~ao Paulo State University (UNESP), S\~ao Paulo, Brazil; E-mail: reilya@ift.unesp.br}}
\footnotetext{\textit{$^{d}$~Division of Materials Theory, Department of Physics and Astronomy, Uppsala University, SE-751 20 Uppsala, Sweden: E-mail: ralph.scheicher@physics.uu.se}}
\footnotetext{\textit{$^{e}$~Federal Institute of Education, Science and Technology of Esp\'{\i}rito Santo, Ibatiba/ES, Brazil: E-mail: fabio.souza@ifes.edu.br}}
\footnotetext{\textit{$^{f}$~Departamento de F\'{\i}sica, Universidade Federal do Esp\'{\i}rito Santo-UFES, Vit\'oria/ES, Brazil; E-mail: wanderla.scopel@ufes.br}}
\footnotetext{\textit{$^{g}$~Departamento de F\'isica, ICEx, Universidade Federal Fluminense-UFF, Volta Redonda/RJ, Brazil; E-mail: rgamorim@id.uff.br}}
\footnotetext{\textit{$^{h}$~Integrated Nanotechnology Research Center, Department of Physics, Faculty of Science, Khon Kaen University, Khon Kaen, 40002, Thailand; E-mail: vittaya@kku.ac.th}}

\footnotetext{\dag~Electronic Supplementary Information (ESI) available: [details of any supplementary information available should be included here]. See DOI: 10.1039/b000000x/}


\section{Introduction}
The development of DNA sequencing tools has advanced in strides over the past decades.\cite{dna_sequencing_review,ngdna_sequencing} Nonetheless the cost for sequencing a genome is yet too expensive for widespread use in personalized medicine\cite{Wheeler:2008bb,Ziegler:2012hn,Shendure:2017bu}. Nanopores have been heralded as a possible way of providing low-cost whole-genome sequencing by measuring either ionic currents as DNA translocates through a nanopore on a membrane,\cite{Dekker:2007tg,doi:10.1021/nl051063o,doi:10.1021/nl051199m} or transverse electronic currents across the membrane itself\cite{B813796J,Howorka:2001es}. A number of  materials have been proposed as scaffolds for this type of molecular recognition, including lipid membranes,\cite{lipid_membrane_and_review} and more recently two dimensional (2D)  solid-state materials such as dichalcogenides,\cite{dichalcogenides,mos2_pores} and specially graphene nanopores.\cite{graph_nanopore_exp_1,de2017electrical,graph_nanopore_teo_1,graph_nanopore_teo_2,nanopore_review_1,nanopore_review_2,nanopore_review_3} While 2D materials represent the best possible surface to volume ratios - a key ingredient in detection - single molecule recognition remains challenging, specially if one is looking for all-electronic methods.\cite{Feliciano:2015ey,doi:10.1021/acs.jpcb.7b03475,nature_1,nature_2}  Graphene combines the feature of being the thinnest possible membrane with very strong carbon-carbon bonds.\cite{graphene_1,graphene_2} However, low signal-to-noise ratios in nanopore experiments have hindered its use in DNA sequencing.\cite{Drndic:2014gh} As an alternative, extended line defects, formed by a combination of octagons and pentagons (OP), have been recently synthesized.\cite{lahiri2010extended,chen2014controlled} These defects have been predicted to act as conducting quasi-one-dimensional wires embedded in the graphene matrix.\cite{lahiri2010extended,AmorimCarbon} Indeed, reducing the dimensionality is a possible path towards improving sensitivity. An added advantage of this arrangement is that the  overall stability of graphene is maintained while allowing for atomically precise edges .

In this work we introduce, extended line defects in graphene to improve the performance of graphene-based DNA sensors. We use a combination of quantum mechanics/molecular mechanics (QM/MM) hybrid methods,  and non-equilibrium Green's functions  (NEGF)\cite{datta,Brandbyge2002,Rocha:2005ep,PhysRevB.73.085414} to investigate the electronic conductance modulation along extended line defects connecting a nanopore explicitly including the effects of the solvent \cite{Feliciano:2015ey,doi:10.1021/acs.jpcb.7b03475}. By sampling over a large number of different orientations, we theoretically demonstrate that distinction between four nucleobases can be achieved.   We further show that the changes in transport are associated with electron flow across specific molecular states near the Fermi level, and their respective coupling to the pore.

\section{Computational Methods}

\begin{figure}[ht!]
\center
\includegraphics[width = 8.5 cm]{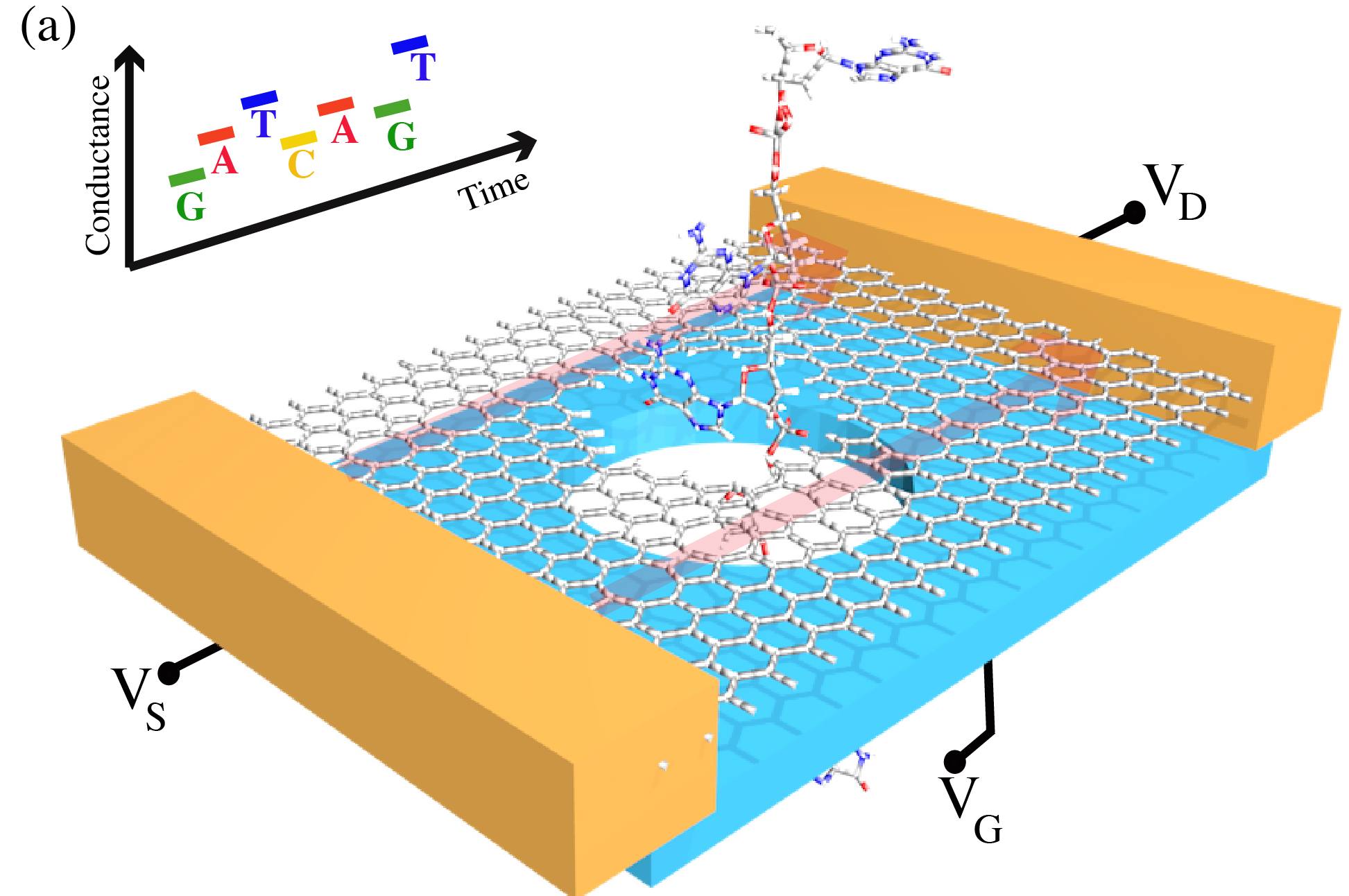}\\
\vspace{0.5cm}
\includegraphics[width = 8.5 cm]{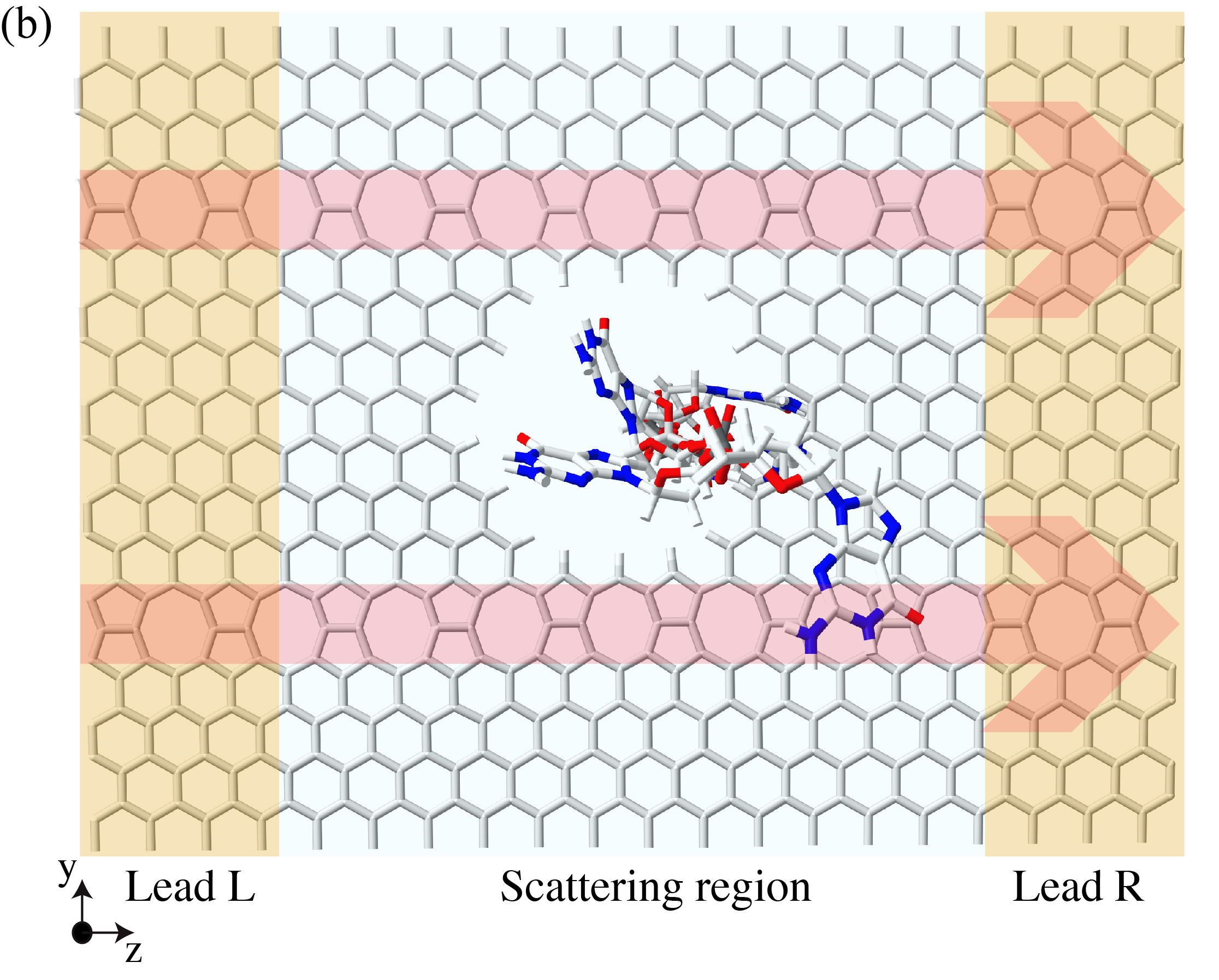}
\caption{a) Schematic setup for a DNA sequencing device  constructed by embedding line defects around a nanopore in graphene connected to a source and drain (in yellow color), with an additional gate electrode (in blue color) below. The transverse conductance would be measured as a DNA strand passes through the nanopore. The different conductance signals for each nucleobase provide a time-resolved signature for the DNA sequence.
b) The device setup used for our first-principles quantum transport simulation consists of a scattering region and two semi-infinite leads. In the scattering region, a nanopore is created between two line defects in graphene, whose edge carbon atoms are passivated by H, and the nucleotides are inserted into the pore. When a solvent is present we also include a buffer layer at the interface between the leads and the scattering region, where the solvent potential drops to zero.}\label{TRCsetup}\label{fig:transp_setup}
\end{figure}
 
The prototypical sequencing device is conceived as shown in Figure \ref{TRCsetup}a.  It consists of a graphene sheet containing two parallel line defects. Along the ribbon formed between the topological defects we introduce a nanopore, created at a location where its edges are close to both line defects. If a source-drain bias voltage is applied along the device - in the direction of the line defects - as DNA is translocated through the pore, changes in conductance during that process will indicate the type of basis residing inside the pore at any given time. Finally, application of a gate voltage introduces an extra degree of freedom to tune the chemical potential to a region which provides maximum sensitivity.

The specific setup for our calculations is shown in figure \ref{TRCsetup}b. It consists of a graphene sheet containing two parallel OP line defects extending along the entire system, and a hydrogen-saturated nanopore with a diameter of $\sim 12$ \AA. The graphene is periodic in the x-direction, while in the z-direction, it is attached to semi-infinite periodic structures (one unit cell of electrodes is shown in blue in Figure \ref{TRCsetup}b).  For our calculations, we consider a single nucleotide at a time inside the pore, as we have already shown that the effects of adjacent bases on sensing is small\cite{Feliciano:2015ey,doi:10.1021/acs.jpcb.7b03475}. The four nucleotides; deoxyadenosine monophosphate (dAMP), deoxythymidine monophosphate (dAMT), deoxycytidine monophosphate
(dAMC) and deoxyguanosine
monophosphate (dAMG), are abbreviated simply as A, T, C and G, respectively. 

First we performed a full dynamical treatment for the four nucleotides (and also the empty pore) in solution (water and 0.2 M of Na and Cl) using classical molecular dynamics. 
These were carried out in two steps: an initial NPT simulation (Berendsen barostat and Noose-Hoover thermostat) to equilibrate the pressure, followed by an NVT production run simulation (with the same thermostat) to obtain the structures. Both MD simulations were carried out using a time-step of 2 fs to generate 20 ns of trajectories. The whole graphene sheet was kept fixed during the simulation and a harmonic potential was applied to a central atom in the nucleotide to keep it inside the pore. All MD calculations were carried out using GROMACS,\cite{gromacs} with the AMBER99SB\cite{amber} force field and single-point-charge (SPC) water model\cite{spc1,spc2}.

\begin{figure*}[ht!]
\center
\includegraphics[width = 18.6 cm]{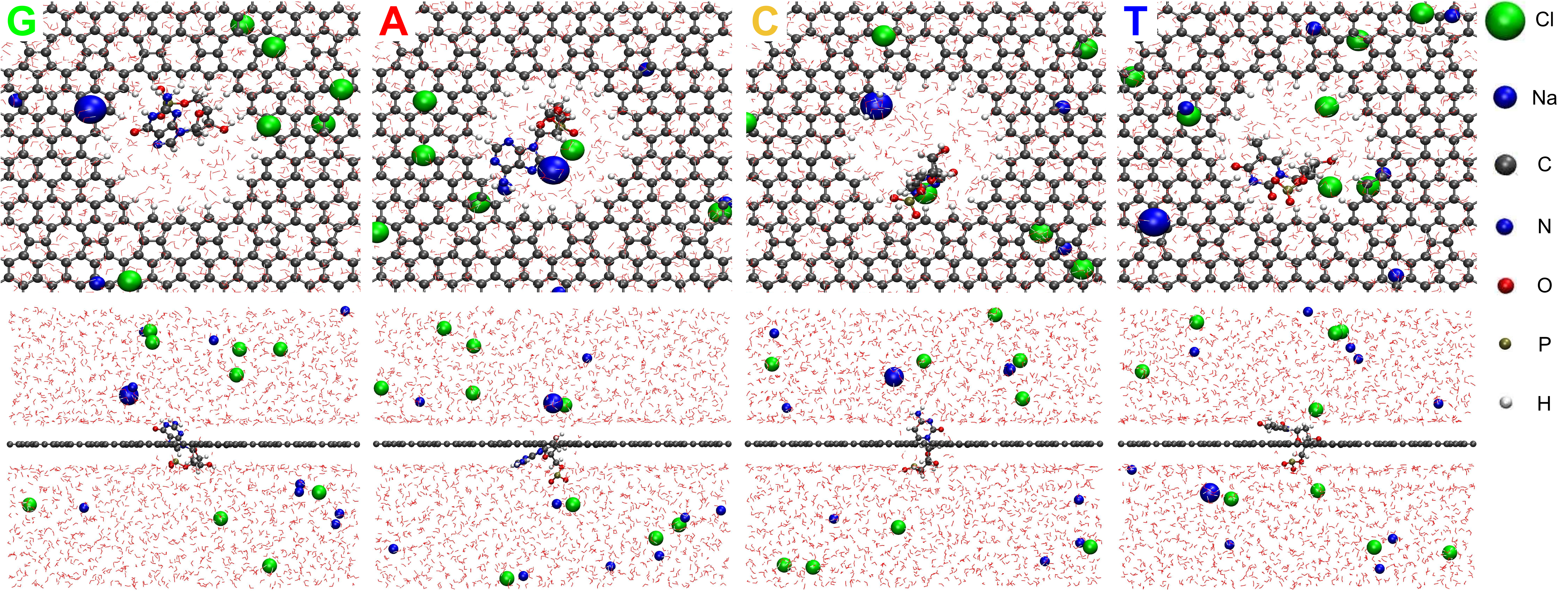}
\caption{Top view (upper panels) and side view (lower panels) for characteristic frames from the molecular dynamics simulations for G, A, C, and T inside the nanopore. In the QM/MM simulation, graphene, the nucleotide and one Na atom make up the QM part of the system whereas the remaining atoms are considered classically (MM). The Na ion considered as QM is always the one closest to the nucleotide, and is shown with a larger radius.}\label{fig:A_T_C_G_systems-in-water}
\end{figure*}

From the NVT simulation we extracted 50 snapshots equally separated in time. Typical configurations for each of the nucleobases in the graphene nanopore are shown in Figure \ref{fig:A_T_C_G_systems-in-water}.  We then  divided the system into a QM and an MM region to perform the QM/MM simulation\cite{qmmm_1,ursuladna}. In our case the QM part includes graphene, the nucleotide and one Na atom to equilibrate the charge. All water molecules and the remaining Na and Cl ions were chosen as the MM part.

The QM part of the calculation is performed within DFT as implemented in SIESTA,\cite{Soler2002a}  modified to include the MM contribution as a static external potential. We employed the generalized gradient approximation (GGA) to the exchange-correlation functional,\cite{Perdew1996a} and valence electrons are described using a local basis set with a single-$\zeta$ plus polarization orbital (SZP) for all elements, which has already been proven a valid approach for the nucleobase-graphene system \cite{doi:10.1021/nl200147x,doi:10.1021/jp4048743}. The atomic core electrons are modeled using Troullier-Martins norm-conserving pseudopotentials.\cite{Troullier1991} For the k-point mesh generation, $5 \times 1 \times 5~ k$-points were employed and the mesh cutoff energy was 150 Ry. 

The DFT calculation is combined with the NEGF technique to determine the electronic transport.\cite{datta,Soler2002a,PhysRevB.73.085414} The system is divided into three regions, namely a scattering region and two semi-infinite electrodes as shown in Figure \ref{TRCsetup}. We self-consistently calculate the charge density using the Green's function for the scattering region, including the effects of the electrodes via self-energies and assuming the Kohn-Sham Hamiltonian as a single-particle Hamiltonian for our problem, as implemented in the SMEAGOL package \cite{Rocha:2005ep,PhysRevB.73.085414}. In our particular case the Hamiltonian incorporates the effects of the solvent from our QM/MM calculation. The exchange-correlation functional and basis sets employed in the transport calculation are identical to those described above for the electronic structure calculations. We considered only the $\Gamma$-point for Brillouin zone sampling. Within the NEGF approach, the quantity we are interested in is the transmission coefficient, 
\begin{equation}
T(E)=\mbox{Tr}[\Gamma_R(E)G^R(E)\Gamma_L(E)G^A(E)] ~,
\end{equation}
where $\Gamma_{L,(R)}(E)$ is the broadening matrix of the left (right) electrode, and $G^{R,(A)}(E)$ is the retarded (advanced) Green's function. Then, we simply evaluate the conductance, $G=G_0T(E_F)$, where $G_0=2e^2/h$ is the quantum conductance, $e$ is the charge of the electron and $h$ is Planck's constant. A more detailed description of the methodology is presented elsewhere\cite{datta,Soler2002a,PhysRevB.73.085414} including the QM/MM transport method \cite{Feliciano:2015ey,doi:10.1021/acs.jpcb.7b03475}.

\section{Results}

\begin{figure}[h!]
\includegraphics[width = 8.5 cm]{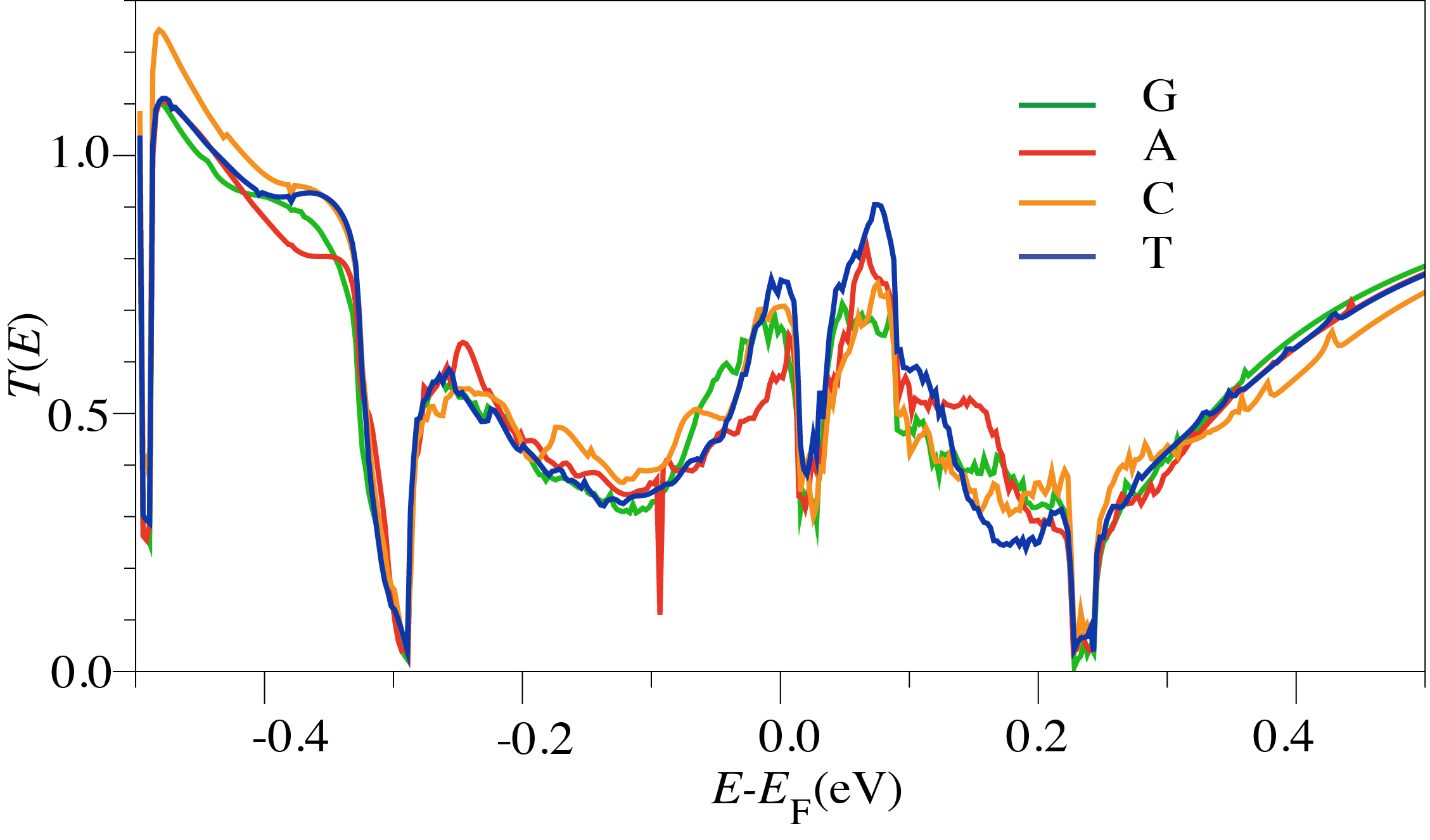}
\caption{Average total transmission coefficients as a function of energy for the four different nucleotides (G, A, C, T). The averages were taken over 50 snapshots from the MD simulations.}\label{fig:G_A_C_T_L_average-transp}
\end{figure}

Figure \ref{fig:G_A_C_T_L_average-transp} shows  the transmission averaged over 50 electronic transport calculations for uncorrelated structural configurations of each system. We have also performed calculations for an empty pore in a solvent (see Supplementary Information). For energies close to the Fermi energy, the transmission for the empty pore in liquid is smaller than the transmission for the nucleotides.  In fact, we notice that the effect of introducing any of the basis into the pore is an overall shift of the transmission due to the interaction with the nucleobase. As a result the sensitivity of the pore to any nucleobase is expected to be very high.   
Moreover, we can also see that the four nucleotides present different average transport properties for energies slightly above the Fermi energy  (up to 0.1 eV). 

\begin{figure}[ht!]
	\centering
	\includegraphics[width = 8.5 cm]{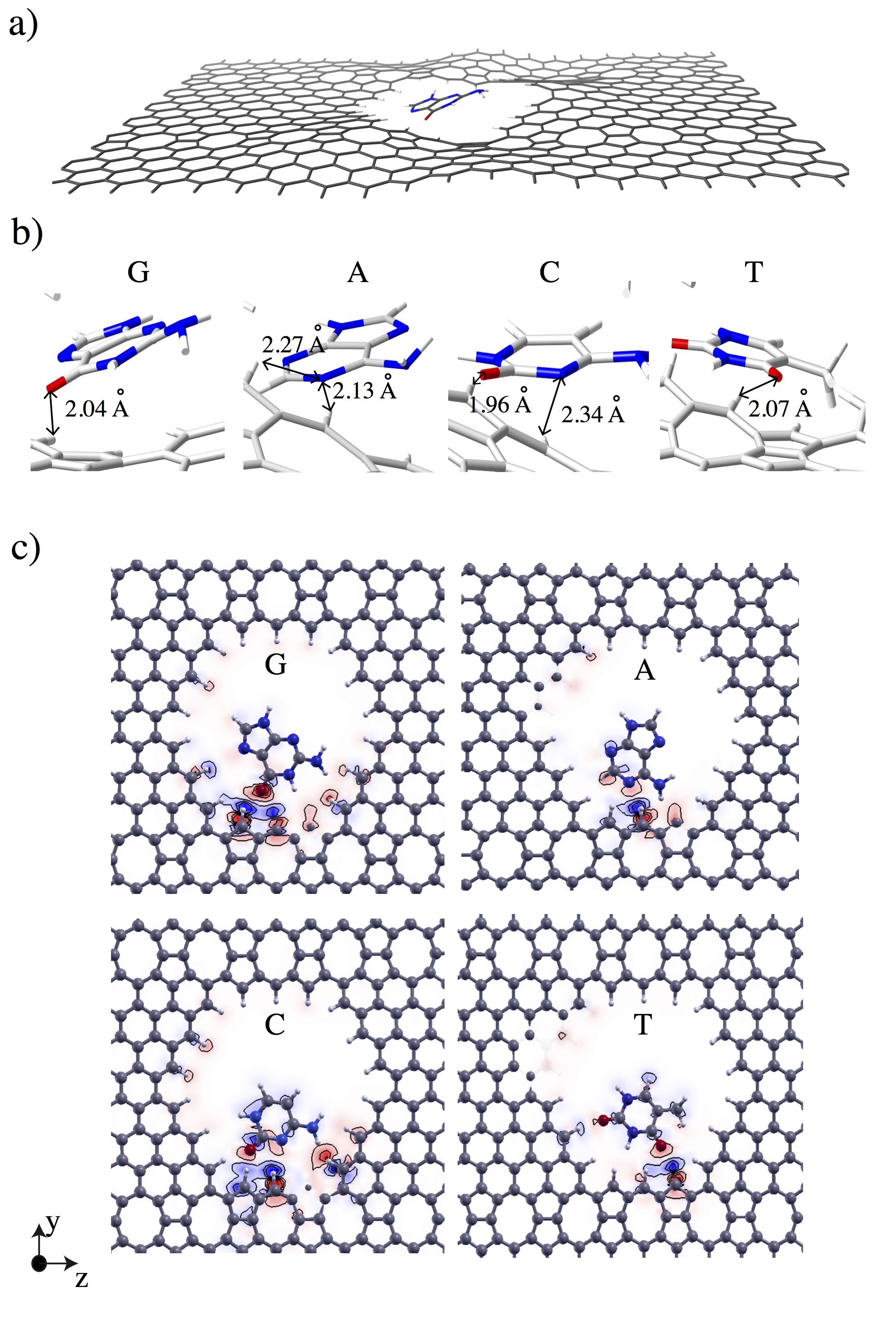}
	\caption{a) The fully optimized structure of the device setup used in the DFT calculations. The wavy structure of graphene is caused by binding between the base and the line defects. b) The most stable configurations for the four nucleobases G, A, C, T. c) Charge density differences (isovalue $0.002 \ e / \text{\AA}^3$ in the y-z graphene plane) of G, A, C, T bound to one side of the line-defect graphene. These binding sites yield the largest binding energies. Blue and red colors represent charge depletion and accumulation, respectively.}\label{graphene}
\end{figure}

\begin{figure}[t]
	\centering
	\includegraphics[width = 9 cm]{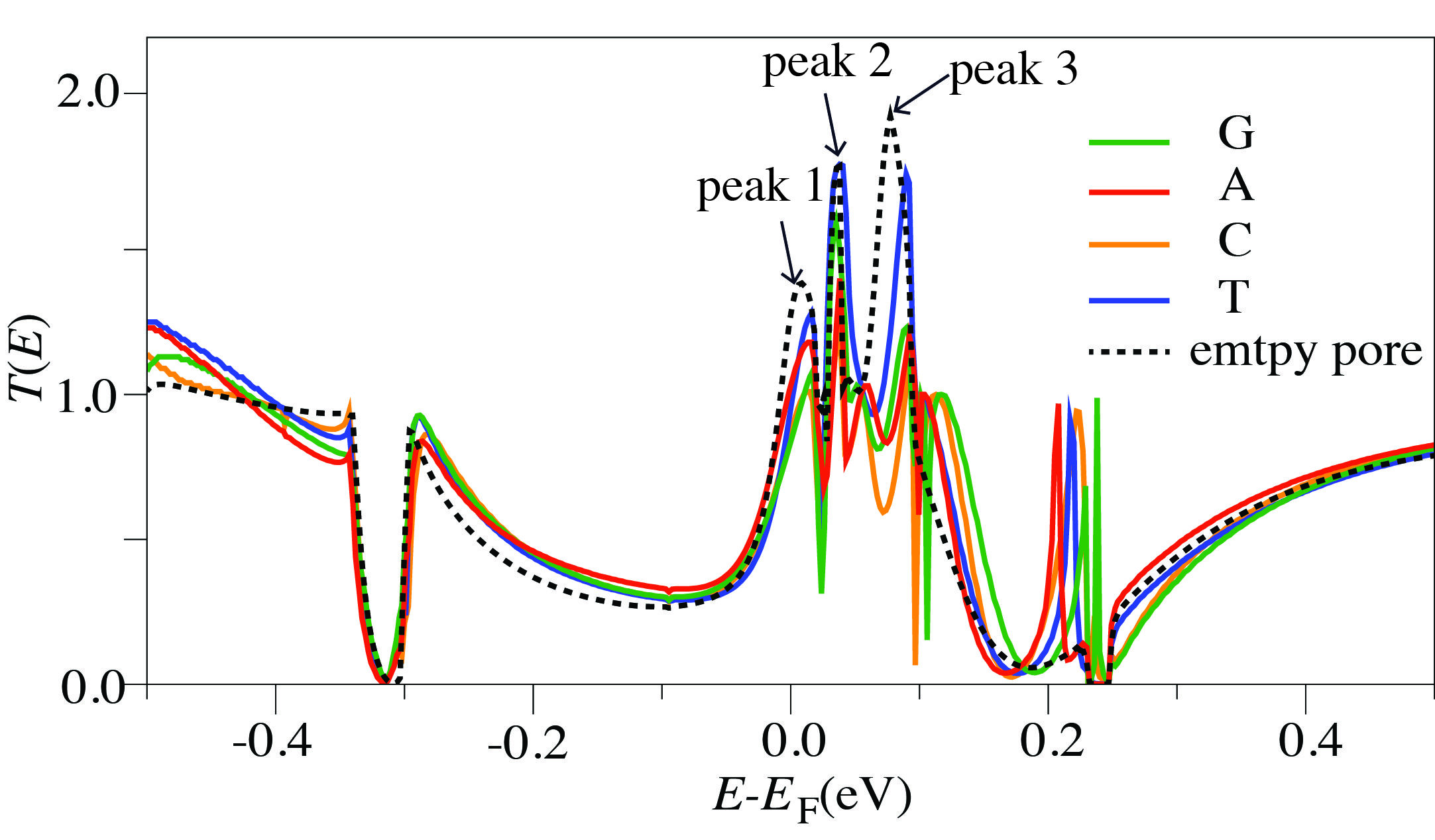}
	\caption{Zero-bias transmission coefficients as a function of energy responsible for the most favorable configuration shown in Figure \ref{graphene}c, as compared to that in the absence of nucleobase.}\label{trc}
\end{figure}

In order to understand the origin of the changes in conductance, and to determine trends in transport, and identify potential descriptors, we performed DFT calculations of the isolated nucleobases in the graphene nanopore. All nucleobases were initially positioned so as to lie in the plane of the graphene. After full relaxation, the interaction between nucleobases and line defects induce a wavy structure  on the graphene sheet. A typical final arrangement for a base within the pore is presented in Figure \ref{graphene}a 
for the case of guanine. Note that the sugar-phosphate backbone, counterions and water molecules were excluded from the quantum transport calculations \cite{Prasongkit:2015ia}. Fig \ref{graphene}b  shows favorable binding configurations of four nucleobases and identifies bond lengths between the bases and graphene,  which are smaller  than 2.5 \AA.  The binding energy $E_{\mbox{b}}$ between the nanopore and nucleobase was calculated  as
\begin{equation}
E_{\mbox{b}}= (E_{\mbox{nanopore}}+E_{\mbox{base}})-E_{\mbox{total}} ~,
\end{equation} 
where $E_{\mbox{total}}$ is the total energy of the system (graphene nanopore+base), and $E_{\mbox{nanopore}}$ and $E_{\mbox{base}}$ is the energy of graphene nanopore and nucleobase, respectively. The calculated binding energies ($E_b$) of the most favorable configurations shown in Figure \ref{graphene} are found to be 0.30, 0.33, 0.41 and 0.23 eV for G, A, C and T, respectively.

These binding energies are due to the charge-transfer interactions between the non-bonding electrons (oxygen, nitrogen) and  hydrogen. The interaction between oxygen and hydrogen bond is stronger than that of nitrogen and hydrogen since oxygen has a higher electronegativity. However, the binding energy of A is  slightly higher than that of G, since A uses its non-bonding electrons (N) to interact with 2 hydrogen atoms. For 
C, the high binding energy is due to two interacting sites forming H-bonds (O-H and N-H). Finally, compared to G, T has a smaller electronic density around the oxygen atom bound to the graphene  due to the presence of a second oxygen, therefore, it has lower binding energy. 

\begin{figure}[t]
	\centering
	\includegraphics[width = 9.8 cm]{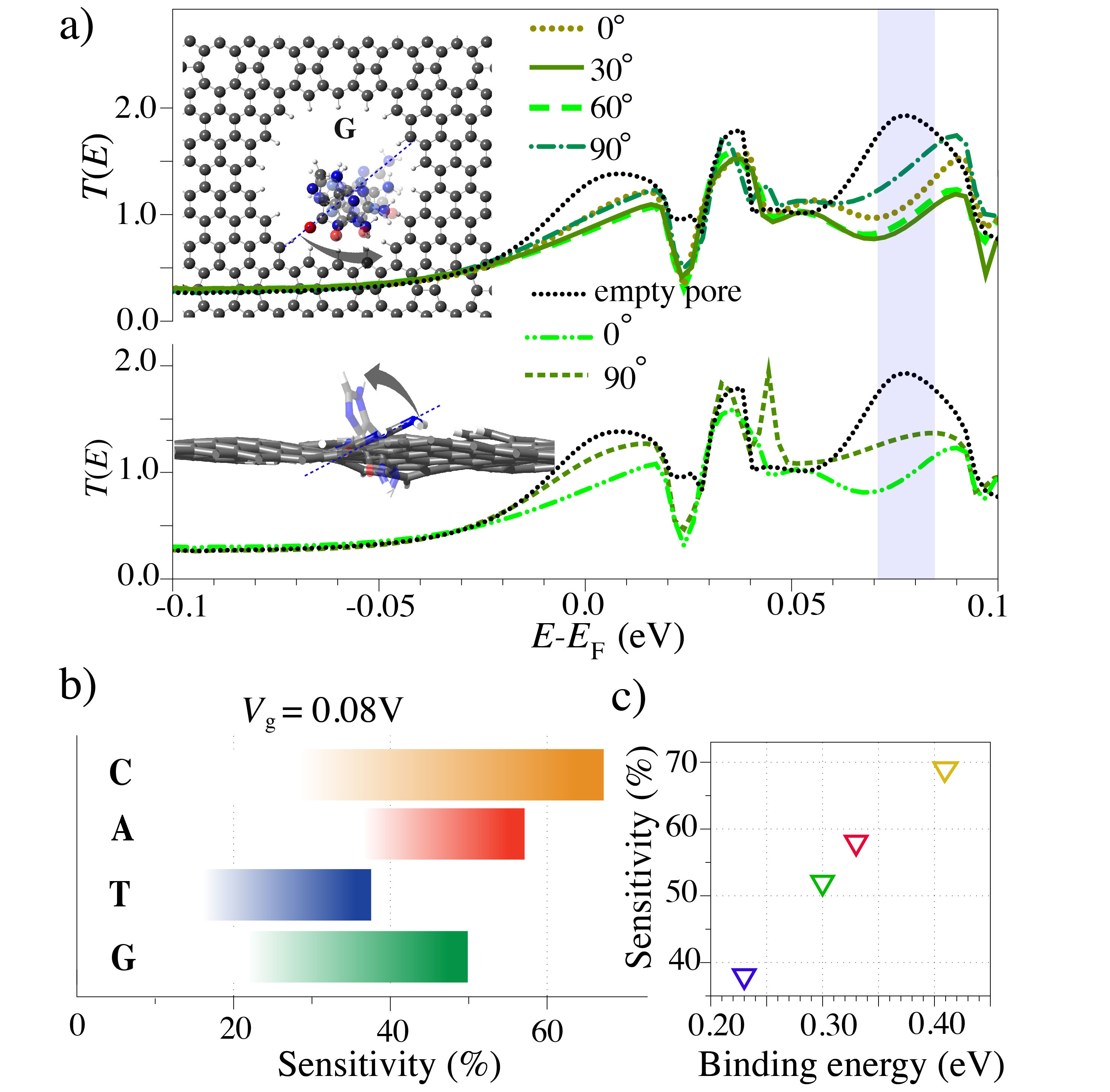}
	\caption{a) The transmission coefficients as a function of energy in which a G nucleotide is rotated in the y-z (top panel) and x-z plane (bottom panel), respectively. b) The variation in the sensitivity at $V_g$=0.08 V is due to the variation of molecular orientations for C, A, T, G bases. The color scale from light to dark refers to the binding strength between the bases and the nanopore edges. c) The sensitivity vs. binding energy of the four nucleobases whose configurations and transmission coefficients are illustrated in Figure \ref{graphene}b and Figure \ref{trc}, respectively.}\label{rotate}
\end{figure}

We have mapped the total charge density redistribution due to the interaction between device and nucleobase by comparing total charge of the system with its isolated parts: nucleobase and device, using the expression: $\Delta \rho(\vec{r}) = \rho_{device + base}(\vec{r})-(\rho_{base}(\vec{r})+\rho_{device}(\vec{r}))$. The charge density difference plotted in Figure \ref{graphene}c reveals that charge fluctuates mostly around the line defect close to the bases. This charge reorganization causes the modulation of charge transport in extended line defects. In Figure \ref{trc}, we present the zero-bias transmission coefficients as a function of energy responsible for the most favorable configuration shown in Figure \ref{graphene}b. By looking at the transmission coefficients of  the empty pore, there are three resonance peaks lying close to the Fermi energy for an energy range  $E-E_F\in \left[0.0, 0.1\right]$ eV. Peaks 2 and 3 are sharper and taller since both are responsible for the pathway of the wavefunction flowing through the extended line defect (see Supplementary Information of electrodes for details).
As seen in Figure \ref{trc}, the overall transport properties for this dry system are qualitatively similar to the case with solvent, which allows us to identify the main features that modulate the electronic transport.
When the nucleobases are inserted into the nanopore, the height of transmission peak 1, 2 and 3 decrease due to interaction between the base and graphene. Especially for peak 3 (at $E-E_F \sim 0.08$ eV)  for which two conducting channels are available, the peak height is very sensitive to the interaction strength. Therefore, we selected orientations of  the four nucleobases in order to analyze the effect of binding configurations on the transmission function. An illustration of the rotation of guanine and its transmission as a function of the angle is shown in Figure \ref{rotate}a. 
 
By rotating in the y-z and x-z plane (see Figure \ref{rotate}a), the transport properties can be modulated. Note that each structure is fully relaxed again by using DFT for the final configurations. It is seen that the specific position of oxygen with respect to line defects strongly affects the transmission  peak height; the stronger  the interaction between the base and graphene, the  greater the  reduction in peak height. Since the pronounced distinction in $T(E)$ is observed at $E-E_F \sim  0.08$ eV in which the current will flow predominantly along the line defect, we assume a positive gate voltage  applied to the device that shifts the Fermi level accordingly.

The sensitivity is assessed by the change in conductance; $S= |(G-G_{0})/G_{0}|\times100\%$, where $G$ and $G_{0}$ are the conductance of the nanopore+base and the empty nanopore, respectively.
The sensitivity variations for four nucleobases are presented in Figure \ref{rotate}b. The maximum conductance follows the hierarchy: C > A > G > T. The change in tone, from dark to light, indicates high to low  sensitivity for each nucleobase. In Figure \ref{rotate}c, we observe a linear relationship between the sensitivity and the binding energy, confirming that the interaction strength significantly affects the sensitivity\cite{Prasongkit:2015ia}. In the experiment, manipulating the translocation speed is critically important for nanopore sequencing \cite{doi:10.1021/nn3051677,Derrington14092010}. The translocation time strongly depends on the orientation of bases \cite{PhysRevE.78.061911,Mathé30082005,Abdolvahab:2008hi}, relating to the binding energy.

Based on our results, we thus note that the existence of line defects in graphene brings about pronounced improvement. By using first-principle investigations, we find that there is an opening of a new conducting channel due to the existence of a metallic wire embedded in a graphene sheet, resulting in higher conductance compared to using only a graphene nanopore. Although the influence of the solvent on the dynamics reduces the conductance by about $50 \%$, there is a significant improvement in selectivity. 

\section{Conclusions}
In conclusion, we have demonstrated that using extended line defects in graphene may present advantages over the most conventional choice based purely on graphene nanopores. The conductance modulation fully accounting for solvent effects has been studied by us using a combination of QM/MM and NEGF methods. Based on DFT formalism, we have investigated the interaction strength between line defects in graphene and nucleobases related to the selectivity and sensitivity of the devices. Furthermore, we have demonstrated the possibility to improve sensing performance by applying a gate voltage to the device. We are confident that our results will encourage future studies to fabricate the proposed device, as well as follow-up theoretical investigations in a search for higher sensing performance depending on pore edge termination by different functional groups.

\section*{Acknowledgements}
The authors acknowledge support from the Thailand Research Fund (MRG5980185), FAPESP (Proj. \# 2016/01343-7), FAPES, CAPES, CNPq-Brazil and the National Laboratory for Scientific Computing (LNCC/MCTI, Brazil) for providing HPC resources of the SDumont supercomputer, which have contributed to the research results reported within this paper. \url{http://sdumont.lncc.br}. ARR acknowledges support from the ICTP-Simmons Foundation associate scheme. RHS thanks the Swedish Research Council (VR) for financial support. The Swedish National Infrastructure for Computing (SNIC) provided computing time for this project. 

\bibliographystyle{rsc} 

\begin{thebibliography}{9}
\bibitem[Heerema and Dekker(2016)]{dna_sequencing_review}
S.~J. Heerema and C.~Dekker, \emph{Nat. Nanotechnol.}, 2016, \textbf{11}, 127 --
136.\relax

\bibitem[Bahassi and Stambrook(2014)]{ngdna_sequencing}
E.~M. Bahassi and P.~J. Stambrook, \emph{Mutagenesis}, 2014, \textbf{29},
303--310.\relax

\bibitem[Wheeler \emph{et~al.}(2008)Wheeler, Srinivasan, Egholm, Shen, Chen,
McGuire, He, Chen, Makhijani, Roth, Gomes, Tartaro, Niazi, Turcotte, Irzyk,
Lupski, Chinault, Song, Liu, Yuan, Nazareth, Qin, Muzny, Margulies,
Weinstock, Gibbs, and Rothberg]{Wheeler:2008bb}
D.~A. Wheeler, M.~Srinivasan, M.~Egholm, Y.~Shen, L.~Chen, A.~McGuire, W.~He,
Y.-J. Chen, V.~Makhijani, G.~T. Roth, X.~Gomes, K.~Tartaro, F.~Niazi, C.~L.
Turcotte, G.~P. Irzyk, J.~R. Lupski, C.~Chinault, X.-z. Song, Y.~Liu,
Y.~Yuan, L.~Nazareth, X.~Qin, D.~M. Muzny, M.~Margulies, G.~M. Weinstock,
R.~A. Gibbs and J.~M. Rothberg, \emph{Nature}, 2008, \textbf{452},
872--876.\relax

\bibitem[Ziegler \emph{et~al.}(2012)Ziegler, Koch, Krockenberger, and Gro{\ss}hennig]{Ziegler:2012hn}
A.~Ziegler, A.~Koch, K.~Krockenberger and A.~Gro{\ss}hennig, \emph{Hum.
	Genet.}, 2012, \textbf{131}, 1627--1638.\relax

\bibitem[Shendure \emph{et~al.}(2017)Shendure, Balasubramanian, Church, Gilbert, Rogers, Schloss, and Waterston]{Shendure:2017bu}
J.~Shendure, S.~Balasubramanian, G.~M. Church, W.~Gilbert, J.~Rogers, J.~A. Schloss and R.~H. Waterston, \emph{Nature}, 2017, \textbf{550}, 345--353.\relax

\bibitem[Dekker(2007)]{Dekker:2007tg}
C.~Dekker, \emph{Nat. Nanotechnol.}, 2007, \textbf{2}, 209--215.\relax

\bibitem[Fologea \emph{et~al.}(2005)Fologea, Uplinger, Thomas, McNabb, and
Li]{doi:10.1021/nl051063o}
D.~Fologea, J.~Uplinger, B.~Thomas, D.~S. McNabb and J.~Li, \emph{Nano
	Lett.}, 2005, \textbf{5}, 1734--1737.\relax

\bibitem[Fologea \emph{et~al.}(2005)Fologea, Gershow, Ledden, McNabb,
Golovchenko, and Li]{doi:10.1021/nl051199m}
D.~Fologea, M.~Gershow, B.~Ledden, D.~S. McNabb, J.~A. Golovchenko and J.~Li,
\emph{Nano Lett.}, 2005, \textbf{5}, 1905--1909.\relax

\bibitem[Howorka and Siwy(2009)]{B813796J}
S.~Howorka and Z.~Siwy, \emph{Chem. Soc. Rev.}, 2009, \textbf{38},
2360--2384.\relax

\bibitem[Howorka \emph{et~al.}(2001)Howorka, Cheley, and
Bayley]{Howorka:2001es}
S.~Howorka, S.~Cheley and H.~Bayley, \emph{Nat. Biotechnol.}, 2001,
\textbf{19}, 636--639.\relax

\bibitem[Bello \emph{et~al.}(2017)Bello, Kim, Kim, Jeon, and
Shim]{lipid_membrane_and_review}
J.~Bello, Y.-R. Kim, S.~M. Kim, T.-J. Jeon and J.~Shim, \emph{Microchim.
	Acta}, 2017, \textbf{184}, 1883--1897.\relax

\bibitem[Loo \emph{et~al.}({2014})Loo, Bonanni, Ambrosi, and
Pumera]{dichalcogenides}
A.~H. Loo, A.~Bonanni, A.~Ambrosi and M.~Pumera, \emph{{Nanoscale}}, {2014},
\textbf{{6}}, {11971--11975}.\relax

\bibitem[Feng \emph{et~al.}({2015})Feng, Liu, Bulushev, Khlybov, Dumcenco, Kis,
and Radenovic]{mos2_pores}
J.~Feng, K.~Liu, R.~D. Bulushev, S.~Khlybov, D.~Dumcenco, A.~Kis and
A.~Radenovic, \emph{{Nat. Nanotechnol.}}, {2015}, \textbf{{10}},
{1070--1076}.\relax

\bibitem[Chen \emph{et~al.}({2017})Chen, Liu, Ouyang, Gao, Liu, and
Zhao]{graph_nanopore_exp_1}
W.~Chen, G.-C. Liu, J.~Ouyang, M.-J. Gao, B.~Liu and Y.-D. Zhao, \emph{{Sci.
		China Chem.}}, {2017}, \textbf{{60}}, {721--729}.\relax

\bibitem[de~Souza \emph{et~al.}(2017)de~Souza, Amorim, Scopel, and
Scheicher]{de2017electrical}
F.~A. de~Souza, R.~G. Amorim, W.~L. Scopel and R.~H. Scheicher,
\emph{Nanoscale}, 2017, \textbf{9}, {2207--2212}.\relax

\bibitem[Farimani \emph{et~al.}({2017})Farimani, Dibaeinia, and
Aluru]{graph_nanopore_teo_1}
A.~B. Farimani, P.~Dibaeinia and N.~R. Aluru, \emph{{ ACS Appl. Mater. Interfaces}}, {2017}, \textbf{{9}}, {92--100}.\relax

\bibitem[Li \emph{et~al.}({2017})Li, Wang, Li, and Han]{graph_nanopore_teo_2}
J.~Li, H.~Wang, Y.~Li and K.~Han, \emph{{Mol Simul.}}, {2017},
\textbf{{43}}, {320--325}.\relax

\bibitem[Abedini-Nassab({2017})]{nanopore_review_1}
R.~Abedini-Nassab, \emph{{Recent Pat. Nanotechnol.}}, {2017},
\textbf{{11}}, {34--41}.\relax

\bibitem[Agah \emph{et~al.}({2016})Agah, Zheng, Pasquali, and
Kolomeisky]{nanopore_review_2}
S.~Agah, M.~Zheng, M.~Pasquali and A.~B. Kolomeisky, \emph{{J. Phys. D Appl. Phys.}}, {2016}, \textbf{{49}}, 413001.\relax

\bibitem[Li \emph{et~al.}(2016)Li, Yu, and Zhao]{nanopore_review_3}
J.~Li, D.~Yu and Q.~Zhao, \emph{Microchimi. Acta}, 2016, \textbf{183},
941--953.\relax

\bibitem[Feliciano \emph{et~al.}(2015)Feliciano, Sanz-Navarro, Coutinho-Neto,
Ordej{\'o}n, Scheicher, and Rocha]{Feliciano:2015ey}
G.~T. Feliciano, C.~Sanz-Navarro, M.~D. Coutinho-Neto, P.~Ordej{\'o}n, R.~H.
Scheicher and A.~R. Rocha, \emph{Phys. Rev. Appl.}, 2015, \textbf{3},
034003.\relax

\bibitem[Feliciano \emph{et~al.}(0)Feliciano, Sanz-Navarro, Coutinho-Neto,
Ordej{\'o}n, Scheicher, and Rocha]{doi:10.1021/acs.jpcb.7b03475}
G.~T. Feliciano, C.~Sanz-Navarro, M.~D. Coutinho-Neto, P.~Ordej{\'o}n, R.~H.
Scheicher and A.~R. Rocha, (in press), \emph{J. Phys. Chem. B}, doi: 10.1021/acs.jpcb.7b03475.\relax

\bibitem[nat(2016)]{nature_1}
Building a better nanopore, \emph{Nat. Nanotechnol.}, 2016, \textbf{11}, 105.\relax

\bibitem[Di~Ventra and Taniguchi(2016)]{nature_2}
M.~Di~Ventra and M.~Taniguchi, \emph{Nat. Nanotechnol.}, 2016, \textbf{11},
117--126.\relax

\bibitem[Novoselov \emph{et~al.}(2004)Novoselov, Geim, Morozov, Jiang, Zhang,
Dubonos, Grigorieva, and Firsov]{graphene_1}
K.~S. Novoselov, A.~K. Geim, S.~V. Morozov, D.~Jiang, Y.~Zhang, S.~V. Dubonos,
I.~V. Grigorieva and A.~A. Firsov, \emph{Science}, 2004, \textbf{306},
666--669.\relax

\bibitem[Geim and Novoselov(2007)]{graphene_2}
A.~K. Geim and K.~S. Novoselov, \emph{Nat. Mater.}, 2007, \textbf{6},
183--191.\relax

\bibitem[Drndi{\'c}(2014)]{Drndic:2014gh}
M.~Drndi{\'c}, \emph{Nat. Nanotechnol.}, 2014, \textbf{9}, 743--743.\relax

\bibitem[Lahiri \emph{et~al.}(2010)Lahiri, Lin, Bozkurt, Oleynik, and
Batzill]{lahiri2010extended}
J.~Lahiri, Y.~Lin, P.~Bozkurt, I.~I. Oleynik and M.~Batzill, \emph{Nat.
	Nanotechnol.}, 2010, \textbf{5}, 326--329.\relax

\bibitem[Chen \emph{et~al.}(2014)Chen, Aut{\`e}s, Alem, Gargiulo, Gautam,
Linck, Kisielowski, Yazyev, Louie, and Zettl]{chen2014controlled}
J.-H. Chen, G.~Aut{\`e}s, N.~Alem, F.~Gargiulo, A.~Gautam, M.~Linck,
C.~Kisielowski, O.~Yazyev, S.~Louie and A.~Zettl, \emph{Phys. Rev. B},
2014, \textbf{89}, 121407.\relax

\bibitem[de~Souza \emph{et~al.}(2017)de~Souza, Amorim, Prasongkit, a~L.~Scopel,
Scheicher, and Rocha]{AmorimCarbon}
F.~A.~L. de~Souza, R.~G. Amorim, J.~Prasongkit, W.~L.~Scopel, R.~H. Scheicher
and A.~R. Rocha, (in press), \emph{Carbon}, 2017.\relax

\bibitem[Datta(2005)]{datta}
S.~Datta, \emph{Quantum Transport: Atom to Transistor}, Cambridge University
Press, 2nd edn., 2005.\relax

\bibitem[Brandbyge \emph{et~al.}(2002)Brandbyge, Mozos, Ordej\'{o}n, Taylor,
and Stokbro]{Brandbyge2002}
M.~Brandbyge, J.-L. Mozos, P.~Ordej\'{o}n, J.~Taylor and K.~Stokbro,
\emph{Phys. Rev. B}, 2002, \textbf{65}, 165401.\relax

\bibitem[Rocha \emph{et~al.}(2005)Rocha, Garc{\'\i}a-Su{\'a}rez, Bailey,
Lambert, Ferrer, and Sanvito]{Rocha:2005ep}
A.~R. Rocha, V.~M. Garc{\'\i}a-Su{\'a}rez, S.~W. Bailey, C.~J. Lambert,
J.~Ferrer and S.~Sanvito, \emph{Nat. Mater.}, 2005, \textbf{4},
335--339.\relax

\bibitem[Rocha \emph{et~al.}(2006)Rocha, Garc\'{\i}a-Su\'arez, Bailey, Lambert,
Ferrer, and Sanvito]{PhysRevB.73.085414}
A.~R. Rocha, V.~M. Garc\'{\i}a-Su\'arez, S.~Bailey, C.~Lambert, J.~Ferrer and
S.~Sanvito, \emph{Phys. Rev. B}, 2006, \textbf{73}, 085414.\relax

\bibitem[Berendsen \emph{et~al.}(1995)Berendsen, van~der Spoel, and van
Drunen]{gromacs}
H.~Berendsen, D.~van~der Spoel and R.~van Drunen, \emph{Comput. Phys. Commun.}, 1995, \textbf{91}, 43 -- 56.\relax

\bibitem[Hornak \emph{et~al.}(2006)Hornak, Abel, Okur, Strockbine, Roitberg,
and Simmerling]{amber}
V.~Hornak, R.~Abel, A.~Okur, B.~Strockbine, A.~Roitberg and C.~Simmerling,
\emph{Proteins: Struct., Funct., Bioinf.}, 2006, \textbf{65},
712--725.\relax

\bibitem[Berendsen \emph{et~al.}(1987)Berendsen, Grigera, and Straatsma]{spc1}
H.~J.~C. Berendsen, J.~R. Grigera and T.~P. Straatsma, \emph{J.
	Phys. Chem.}, 1987, \textbf{91}, 6269--6271.\relax

\bibitem[Berendsen H. J.~C. and J.(1981)]{spc2}
H. J. C. Berendsen, J. P. M. Postma, W. F. van Gunsteren, and J. Hermans, Interaction Models for Water in
Relation to Protein Hydration \emph{Intermolecular Forces, edited by B. Pullman}, Springer, 1981, Vol.~14, 331--342. \relax

\bibitem[Sanz-Navarro \emph{et~al.}(2011)Sanz-Navarro, Grima, Garc{\'i}a, Bea,
Soba, Cela, and Ordej{\'o}n]{qmmm_1}
C.~F. Sanz-Navarro, R.~Grima, A.~Garc{\'i}a, E.~A. Bea, A.~Soba, J.~M. Cela and
P.~Ordej{\'o}n, \emph{Theor. Chem. Acc.}, 2011, \textbf{128},
825--833.\relax

\bibitem[Rothlisberger and Carloni(2006)]{ursuladna}
U.~Rothlisberger and P.~Carloni, Drug-Target Binding Investigated by
	Quantum Mechanical/Molecular Mechanical (QM/MM) Methods, \emph{Lect. Notes Phys.}, 2006, \textbf{704}, 449--779.\relax

\bibitem[Soler \emph{et~al.}(2002)Soler, Artacho, Gale, Garc, Junquera, Ordej,
and Daniel]{Soler2002a}
M.~Soler, E.~Artacho, J.~D. Gale, A.~Garc, J.~Junquera, P.~Ordej and S.~Daniel,
\emph{J. Phys.: Condens. Matter}, 2002, \textbf{2745}, 2745--2779.\relax

\bibitem[Perdew \emph{et~al.}(1996)Perdew, Burke, and Ernzerhof]{Perdew1996a}
J.~P. Perdew, K.~Burke and M.~Ernzerhof, \emph{Phys. Rev. Lett.}, 1996,
\textbf{77}, 3865--3868.\relax

\bibitem[Prasongkit \emph{et~al.}(2011)Prasongkit, Grigoriev, Pathak, Ahuja,
and Scheicher]{doi:10.1021/nl200147x}
J.~Prasongkit, A.~Grigoriev, B.~Pathak, R.~Ahuja and R.~H. Scheicher,
\emph{Nano Lett.}, 2011, \textbf{11}, 1941--1945.\relax

\bibitem[Prasongkit \emph{et~al.}(2013)Prasongkit, Grigoriev, Pathak, Ahuja,
and Scheicher]{doi:10.1021/jp4048743}
J.~Prasongkit, A.~Grigoriev, B.~Pathak, R.~Ahuja and R.~H. Scheicher, \emph{J. Phys. Chem. C}, 2013, \textbf{117}, 15421--15428.\relax

\bibitem[Troullier(1991)]{Troullier1991}
N.~Troullier, \emph{Phys. Rev. B}, 1991, \textbf{43}, 1993--2006.\relax

\bibitem[Prasongkit \emph{et~al.}(2015)Prasongkit, Feliciano, Rocha, He,
Osotchan, Ahuja, and Scheicher]{Prasongkit:2015ia}
J.~Prasongkit, G.~T. Feliciano, A.~R. Rocha, Y.~He, T.~Osotchan, R.~Ahuja and
R.~H. Scheicher, \emph{Sci. Rep.}, 2015,  5, 17560.\relax

\bibitem[Anderson \emph{et~al.}(2013)Anderson, Muthukumar, and
Meller]{doi:10.1021/nn3051677}
B.~N. Anderson, M.~Muthukumar and A.~Meller, \emph{ACS Nano}, 2013, \textbf{7},
1408--1414.\relax

\bibitem[Derrington \emph{et~al.}(2010)Derrington, Butler, Collins, Manrao,
Pavlenok, Niederweis, and Gundlach]{Derrington14092010}
I.~M. Derrington, T.~Z. Butler, M.~D. Collins, E.~Manrao, M.~Pavlenok,
M.~Niederweis and J.~H. Gundlach, \emph{Proc. Natl. Acad. Sci. U.S.A}, 2010, \textbf{107}, 16060--16065.\relax

\bibitem[Luo \emph{et~al.}(2008)Luo, Ala-Nissila, Ying, and
Bhattacharya]{PhysRevE.78.061911}
K.~Luo, T.~Ala-Nissila, S.-C. Ying and A.~Bhattacharya, \emph{Phys. Rev. E},
2008, \textbf{78}, 061911.\relax

\bibitem[Mathé \emph{et~al.}(2005)Mathé, Aksimentiev, Nelson, Schulten, and
Meller]{Mathé30082005}
J.~Math{\'e}, A.~Aksimentiev, D.~R. Nelson, K.~Schulten and A.~Meller,
\emph{Proc. Natl. Acad. Sci. U.S.A}, 2005, \textbf{102}, 12377--12382.\relax

\bibitem[Abdolvahab \emph{et~al.}(2008)Abdolvahab, Roshani, Nourmohammad,
Sahimi, and Tabar]{Abdolvahab:2008hi}
R.~H. Abdolvahab, F.~Roshani, A.~Nourmohammad, M.~Sahimi and M.~R.~R. Tabar,
\emph{J. Chem. Phys.}, 2008, \textbf{129}, 235102.\relax
\end{thebibliography}

\end{document}